\documentclass[sigconf,nonacm]{acmart}
\renewcommand\footnotetextcopyrightpermission[1]{}
\usepackage{booktabs}
\usepackage{tabularx}
\usepackage{amsmath}
\usepackage{tikz}
\usetikzlibrary{positioning, calc, arrows.meta}

\graphicspath{{./}}


\begin{document}

\title{ARGUS: MCP-Grounded Root Cause Analysis for Kubernetes Incidents}

\author{Ergi Senja}
\affiliation{%
  \institution{Department of Computer Science and Engineering (CSE),\\ Chalmers University of Technology and University of Gothenburg}
  \city{Gothenburg}
  \country{Sweden}
}
\email{gussenjer@student.gu.se}

\author{Seyed Mohammad Reza Razavi Zadegan}
\affiliation{%
  \institution{Department of Computer Science and Engineering (CSE),\\ Chalmers University of Technology and University of Gothenburg}
  \city{Gothenburg}
  \country{Sweden}
}
\email{zadegan@student.chalmers.se}

\author{Philipp Leitner}
\affiliation{%
  \institution{Department of Computer Science and Engineering (CSE),\\ Chalmers University of Technology and University of Gothenburg}
  \city{Gothenburg}
  \country{Sweden}
}
\email{philipp.leitner@chalmers.se}

\begin{abstract}
Kubernetes incident triage requires correlating signals from metrics, logs, container state, and messaging systems across multiple monitoring tools, a fragmented workflow that slows diagnosis and contributes to alert fatigue. Large language models (LLMs) have shown promise for automated root cause analysis (RCA), but existing systems rely on custom, system-specific data access layers that cannot be reused across organisations. We present ARGUS, an MCP-grounded RCA assistant that connects a commercial LLM to live Kubernetes observability data through standardised MCP servers covering Kubernetes state, Prometheus metrics, Loki logs, and NATS messaging, and delivers structured diagnostic summaries inside the Slack incident channel where on-call engineers already work. We conduct a preliminary evaluation of ARGUS using three complementary methods: controlled fault injection across ten Kubernetes incident scenarios, rubric-based scoring of the resulting RCA summaries on three dimensions, and semi-structured interviews with six on-call engineers at an industrial partner. ARGUS named the correct root cause in all ten scenarios with an aggregate MCP success ratio of 0.91. Practitioners trusted the diagnostic output but consistently expressed scepticism toward the recommended fixes. Our central finding is a diagnostic/prescriptive asymmetry: ARGUS reliably identifies what went wrong, but is perceived as less reliable or trustworthy at specifying what to do next. This pattern can be observed across all three evaluation methods, and has important implications for future autonomous agentic incident handling systems.
\end{abstract}


\maketitle

\section{Introduction}
\label{sec:intro}

Cloud computing has become the dominant model for modern software deployment~\cite{armbrust2010cloud}, and Kubernetes\footnote{\url{https://kubernetes.io}} has emerged as the industry standard for container orchestration~\cite{burns2016borg}. Organisations rely on it to operate distributed, microservices-based applications at scale. This operational model introduces significant complexity during incident response: when a failure occurs, it tends to propagate across services and manifest simultaneously across metrics, logs, container state, and messaging infrastructure~\cite{pham2024-rca-microservices}. Performing root cause analysis (RCA) under these conditions requires an on-call engineer to correlate diagnostic signals across multiple monitoring tools, typically Prometheus\footnote{\url{https://prometheus.io}} for metrics, Grafana\footnote{\url{https://grafana.com}} for dashboards, Kubernetes for container and workload state, and messaging backends for application-layer signals. The cognitive overhead of this tool fragmentation slows diagnosis and contributes to alert fatigue among the engineers who handle incidents~\cite{google-sre-practical-alerting}.

The field of AIOps~\cite{dang2019aiops} has begun using LLMs to automate parts of this diagnostic process~\cite{aiops}. LLMs can reason through unstructured text, interpret log messages, and correlate signals across diagnostic dimensions without being trained on a specific system. However, a foundational challenge limits their reliability in this setting: the \emph{grounding problem}~\cite{grounding}. Without access to live system state, an LLM cannot distinguish the current incident from patterns encountered during pre-training. It may generate plausible but factually incorrect diagnoses~\cite{Huang2023A}. Existing LLM-based RCA systems address this by connecting models to diagnostic data sources, but their data access layers are custom-built for each deployment, using bespoke wrappers around specific APIs and data formats~\cite{chen2024-rcacopilot,pei2025-flow-of-action}. Adopting any of these systems in a new organisational setting requires re-implementing the entire data access layer.

The Model Context Protocol (MCP), introduced by Anthropic in late 2024, offers a standardised solution to this interoperability challenge~\cite{anthropic-mcp}. MCP defines a client-server protocol in which an LLM agent issues structured requests to MCP servers, which execute the requests and return typed, structured results. This mediates tool access through a well-defined request-response boundary, constraining the model to sanctioned, read-only operations. New data sources can be added by deploying additional MCP servers with appropriate credentials, with no changes to the agent's code or prompt. For operational contexts, this is a direct and practical mechanism for grounding diagnostic agents in live infrastructure data.

We present ARGUS (Agentic Root-cause Guided Unified System), an MCP-grounded RCA assistant for Kubernetes incidents designed and evaluated in collaboration with an industrial partner. ARGUS connects a commercial LLM to live observability data through four read-only MCP servers covering Kubernetes cluster state, Prometheus metrics, Loki\footnote{\url{https://grafana.com/oss/loki/}} logs, and NATS\footnote{\url{https://nats.io}} messaging, and delivers structured diagnostic summaries as Slack\footnote{\url{https://slack.com}} thread replies in the incident channel where engineers already work. It requires no model fine-tuning, no custom API integration per data source, and no changes to the existing alerting or monitoring stack. Its design separates the agent's reasoning from its tool access through a well-defined protocol boundary~\cite{anthropic-mcp}, making write access structurally impossible and new data sources addable by configuration alone. We evaluate ARGUS along three axes: the evidence it retrieves, the quality of its RCA summaries, and how on-call engineers perceive it.

Our contributions are as follows: (1) the design and description of ARGUS, a ChatOps-integrated, read-only RCA assistant that uses MCP as the sole data access mechanism, requires no model fine-tuning, and can be extended to new data sources by deploying additional MCP servers, and (2) a three-method preliminary evaluation. We find that protocol-mediated grounding enables reliable root cause identification across a range of Kubernetes fault types without model fine-tuning, but practitioner trust is consistently asymmetric: engineers rely on the diagnostic output but systematically question the recommended fixes. This points to a concrete design principle for agents: optimise for diagnostic depth first, and treat command-proposing capability as a separate, explicitly authorised function.

This paper is based on a master's thesis project conducted at Chalmers University of Technology and University of Gothenburg, in collaboration with an industrial partner company which elected to remain anonymous.~\cite{argus-thesis} 

\section{Background \& Related Work}
\label{sec:related}

\textbf{Model Context Protocol.} MCP~\cite{anthropic-mcp} is an open standard defining a client-server protocol for LLM tool access. An MCP client (the agent) discovers available tools from a server via a JSON-RPC handshake and invokes them by name with typed arguments. The server executes the request and returns a structured result. The protocol is transport-agnostic, supporting HTTP, WebSocket, and stdio transports. For operational contexts, the separation between the agent's reasoning and the server's execution provides a natural security boundary: the agent can only perform operations that the MCP server explicitly exposes, and a read-only server configuration makes write access structurally impossible regardless of the model's output.

\textbf{Prior approaches to automated RCA.} Earlier automated RCA systems for microservice environments relied on causal discovery and statistical inference over metric time series~\cite{pham2024-rca-microservices}. These methods can surface candidate causal relationships between services without assuming a structure in advance, but they operate on metrics alone, cannot incorporate unstructured signals such as logs and events, and produce graph-structured output rather than natural-language explanations that an on-call engineer can act on under time pressure. LLMs address the heterogeneous-evidence limitation but introduce the grounding problem.

\textbf{LLM-based RCA systems.} RCACopilot~\cite{chen2024-rcacopilot} connects an LLM to incident data through hand-crafted per-incident-type handlers, demonstrating competitive diagnostic accuracy but requiring bespoke engineering per incident type. Flow-of-Action~\cite{pei2025-flow-of-action} guides a multi-agent system with Standard Operating Procedures to reduce hallucination, but both tools and procedures are tied to a specific deployment. MicroRCA-Agent~\cite{tang2025-microrca-agent} fuses logs, traces, and metrics into compact fault features before querying an LLM, using a custom pipeline that cannot be extended to new data sources without reimplementation. SynergyRCA~\cite{xiang2025-synergyrca} builds state and metadata graphs over Kubernetes resources and uses retrieval-augmented generation for RCA, but the graphs are maintained in a separate database with a dedicated pipeline rather than through a reusable tool interface. A commonality between these approaches is that each system solves the data access problem independently: the LLM's diagnostic capability is strong, but the mechanism connecting it to live data is bespoke and non-transferable.

The system most closely related to ARGUS is OpenDerisk~\cite{di2025-openderisk}, which also adopts MCP for a site reliability engineering platform. OpenDerisk is a broad, multi-agent framework covering RCA, risk assessment, change analysis, and testing, with a production deployment processing over 60,000 diagnostic runs per day. ARGUS differs in scope: it is a single ReAct agent scoped to Kubernetes RCA, integrated into the Slack ChatOps workflow that on-call engineers already use, and constrained to read-only diagnosis without causal graph construction or multi-agent orchestration. Where OpenDerisk demonstrates breadth and production scale, ARGUS contributes a narrow, portable design and a multi-method evaluation that makes the agent's failure modes visible in ways that aggregate production metrics cannot.

\textbf{Structured prompting and retrieval.} eARCO~\cite{goel2025-earco} defines an eight-stage diagnostic structure for LLM-based RCA and shows that structured prompts improve accuracy by up to 21\% over unstructured baselines on cloud incident data. Retrieval-augmented generation (RAG)~\cite{lewis2020rag} supplies LLMs with relevant historical context at inference time by encoding past incidents as dense vector embeddings and retrieving semantically similar cases as few-shot examples. Sentence-BERT~\cite{reimers2019-sbert} and FAISS~\cite{johnson2021-faiss} provide the embedding and indexing infrastructure for this retrieval. ARGUS combines both approaches, using the eARCO framework as its prompt structure and a FAISS-backed incident store for few-shot context, while adding MCP-mediated live tool access as the third layer.

\begin{table}[h!]
\centering
\caption{Comparison of representative RCA approaches.}
\label{tab:related}
\footnotesize
\setlength{\tabcolsep}{3pt}
\begin{tabularx}{\linewidth}{@{}p{2.3cm} c c c c c@{}}
\toprule
System & Live data & MCP & K8s & ChatOps & Eval. \\
\midrule
Chen et al.~\cite{chen2024-rcacopilot} & Custom & No & No & No & Large-scale \\
Pei et al.~\cite{pei2025-flow-of-action} & Custom & No & Partial & No & Prototype \\
Xiang et al.~\cite{xiang2025-synergyrca} & Yes & No & Yes & No & Prod.\ data \\
Di et al.~\cite{di2025-openderisk} & Yes & Yes & Partial & No & Prod.\ scale \\
\textbf{ARGUS (this paper)} & \textbf{Yes} & \textbf{Yes} & \textbf{Yes} & \textbf{Yes} & \textbf{Pilot} \\
\bottomrule
\end{tabularx}
\end{table}

Table~\ref{tab:related} summarises positioning across key properties. No prior system combines standardised MCP tool access, a Kubernetes-specific focus, ChatOps delivery, structured prompting over live evidence, and a multi-method evaluation covering evidence retrieval capability, summary quality, and practitioner perception.

\section{ARGUS System Design}
\label{sec:design}

We now discuss the design and implementation of ARGUS.

\subsection{Architecture Overview}
\label{subsec:architecture}

\begin{figure}[h!]
    \centering
    \includegraphics[width=\linewidth]{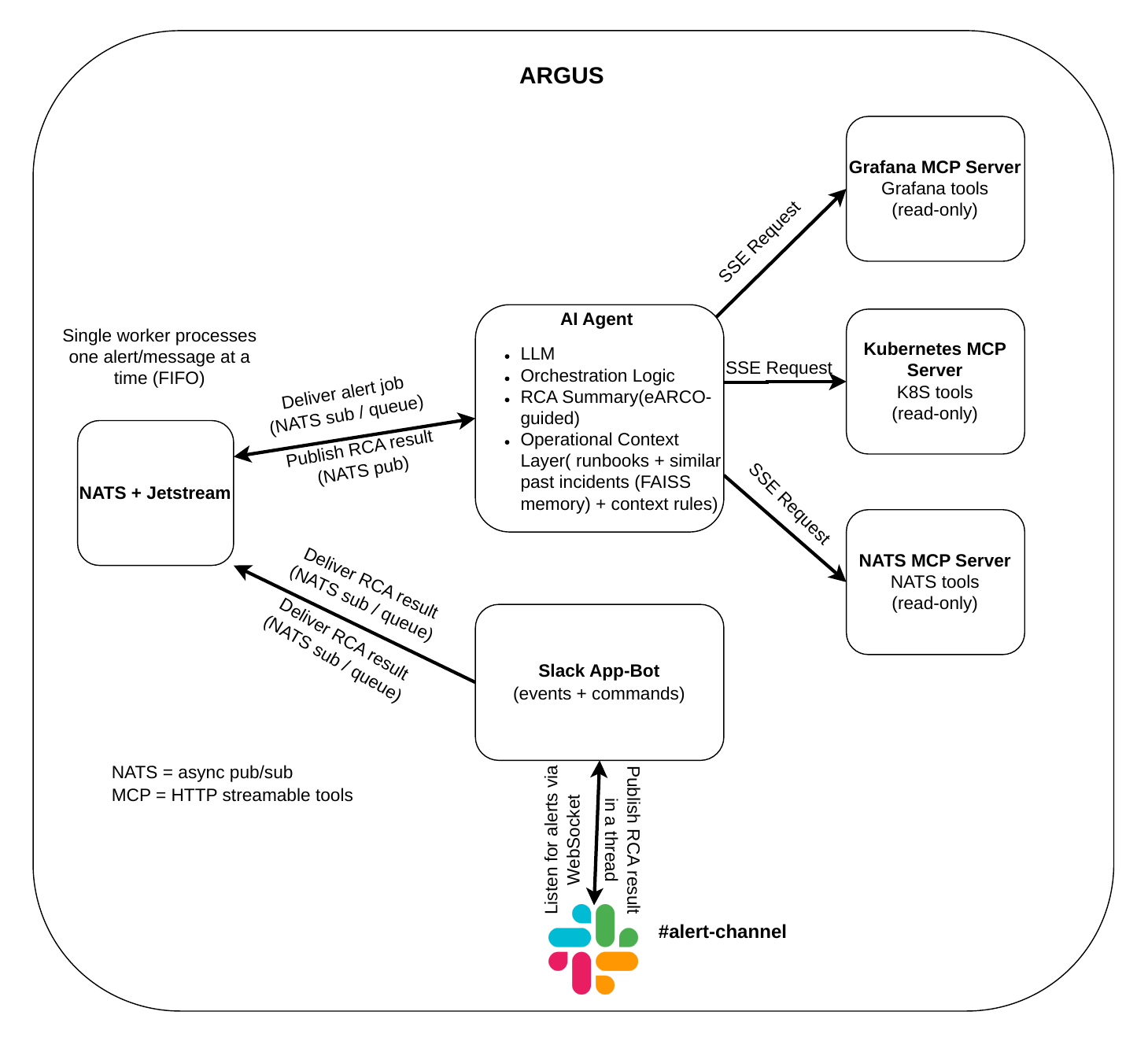}
    \caption{High-level ARGUS architecture. The Slack application and AI agent are decoupled via NATS JetStream. The agent enriches alerts with retrieval-augmented context and then accesses live observability data through four read-only MCP servers before publishing the OperatorBrief.}
    \label{fig:architecture}
    \Description{System architecture diagram showing a Slack application and AI agent connected via NATS JetStream, with four MCP servers (Kubernetes, Grafana/Prometheus, Grafana/Loki, NATS) providing read-only observability data to the agent.}
\end{figure}

ARGUS comprises two application components connected through NATS JetStream as an asynchronous messaging backbone: a Slack application and an AI diagnostic agent. Figure~\ref{fig:architecture} shows the high-level architecture. The Slack application maintains a persistent WebSocket connection to the organisation's incident channel and publishes each incoming Alertmanager\footnote{\url{https://prometheus.io/docs/alerting/latest/alertmanager/}} alert as a job to a NATS JetStream queue. The AI agent subscribes to this queue, processes each alert sequentially through a four-stage pipeline (context retrieval, MCP-based evidence gathering, report generation, and delivery), and publishes the resulting OperatorBrief back to NATS. The Slack application then posts the OperatorBrief as a thread reply to the originating alert message, keeping the diagnosis visible and traceable within the existing incident workflow. The two components are fully decoupled: the Slack application has no knowledge of how the diagnosis is produced, and the agent has no knowledge of how the result is delivered. This separation allows each component to be updated, scaled, or replaced independently.

\subsection{MCP Tool Inventory}
\label{subsec:mcp}

ARGUS exposes observability data through four read-only MCP servers, each covering a distinct layer of the Kubernetes stack. All servers are configured with read-only credentials, making it structurally impossible for the agent to modify cluster state regardless of its reasoning.

The \textbf{Kubernetes MCP server} exposes \texttt{kubectl}-style introspection tools organised into five categories. Pod tools include listing pods by namespace, getting individual pod objects, retrieving current and previous container logs, and reading per-pod CPU and memory usage. Workload tools cover deployment and replica set inspection, job status, and HPA state. Node tools provide node object retrieval, node-level log access, and per-node resource consumption. Event tools list cluster and namespace-scoped events. Resource tools provide generic multi-API-group introspection, allowing the agent to fetch any Kubernetes object by group, version, and kind without a dedicated endpoint. The server runs inside the cluster under a dedicated service account bound to a read-only ClusterRole. The scope of that ClusterRole determines the agent's observability boundary: cluster-scoped resources such as nodes require explicit cluster-level RBAC grants that are absent by default, which reflects the access boundary a least-privilege SRE tool faces in a production environment.

The \textbf{Grafana MCP server} exposes PromQL query execution, LogQL query execution, datasource listing, and label discovery against all Grafana-provisioned datasources. The agent must discover datasource UIDs through a listing call at the start of any metric or log query rather than hardcoding them, making the server resilient to changes in the Grafana configuration. The MCP client manager injects the cluster's default datasource UIDs as server-level argument defaults, reducing the number of discovery calls needed for straightforward queries.

The \textbf{NATS MCP server} exposes read-only introspection over JetStream accounts, streams, consumers, and key-value buckets. It is parameterised with per-account credential files so the agent can query any NATS account for which a credential is provisioned. The credential boundary is the primary access control mechanism: accounts without a provisioned credential are not observable, which produced two failures in the evaluation when the agent attempted to introspect development-environment accounts.

An optional \textbf{GitLab}\footnote{\url{https://gitlab.com}} \textbf{MCP server} is configured but was not activated during the evaluation runs. It illustrates the extensibility of the MCP-based design: adding a new data source requires deploying an MCP server with appropriate credentials and restarting the agent, with no changes to the agent's code or prompt structure.

The MCP client manager routes tool calls from the ReAct loop to the appropriate server. It applies a three-layer argument injection scheme before forwarding each call: server-level defaults fill in values such as datasource UIDs that the LLM should not need to discover repeatedly, per-tool defaults fill any omitted arguments that have a known sensible value, and per-tool overrides unconditionally inject arguments that must always be present. This scheme reduces the number of failed calls caused by missing arguments while preserving the agent's ability to provide specific values when the retrieved context warrants it.

\subsection{Retrieval-Augmented Context}
\label{subsec:rag}

Before the investigation loop begins, ARGUS enriches the alert prompt with three complementary retrieval sources that provide domain-specific context beyond what MCP tool calls can supply at inference time.

\begin{figure*}
    \centering
    \includegraphics[width=\linewidth]{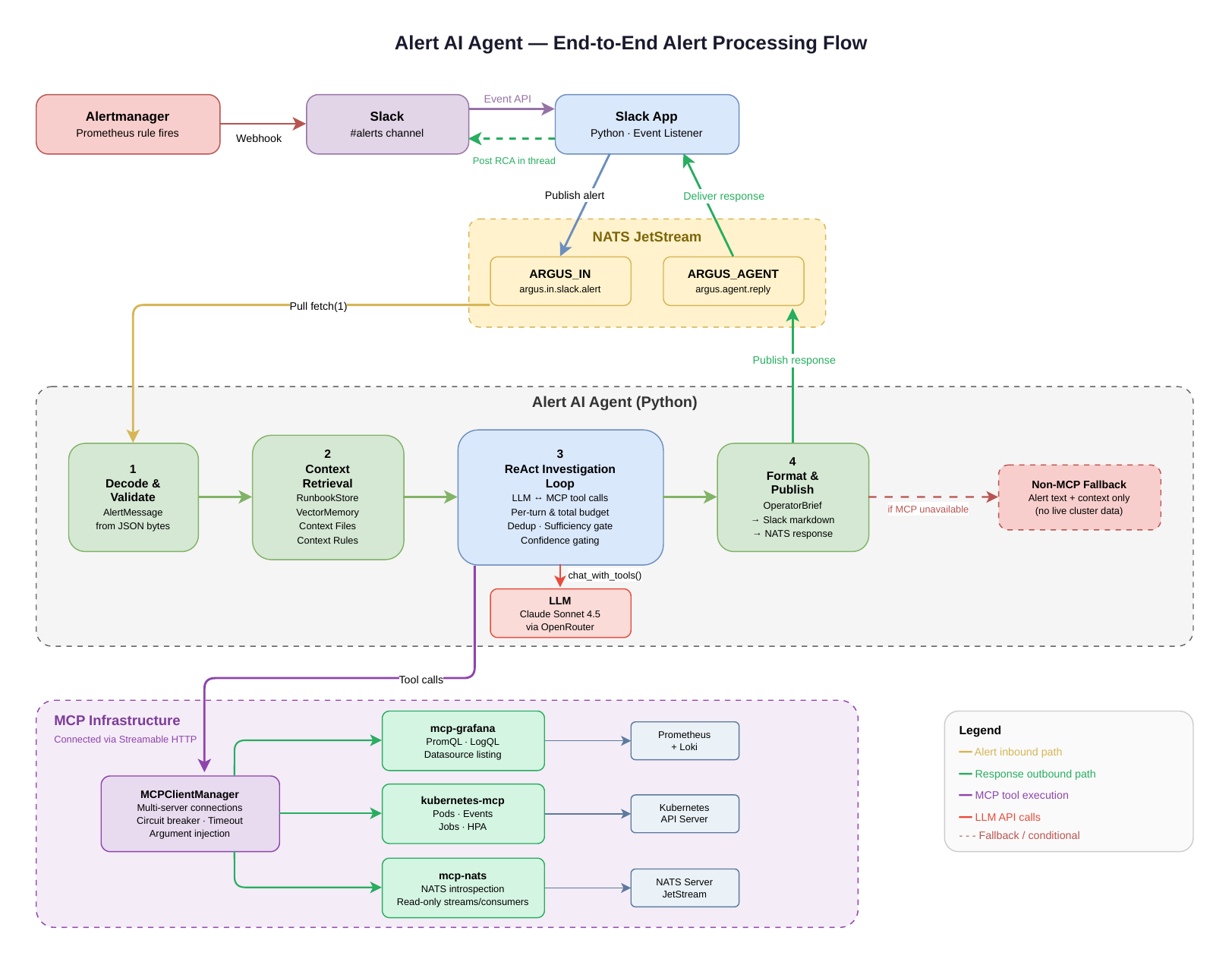}
    \caption{End-to-end alert processing flow in ARGUS. The four agent stages are shown alongside MCP tool execution paths, the deduplication and budget guardrails, and the non-MCP fallback used when servers are unavailable.}
    \label{fig:e2e}
    \Description{Flow diagram showing alert processing stages: context retrieval, MCP evidence gathering with ReAct loop, report generation, and Slack delivery. MCP paths show tool calls to four servers with deduplication and budget guardrails. A fallback path bypasses MCP when no servers are available.}
\end{figure*}

The \textbf{runbook store} provides structured diagnostic procedures for common Kubernetes alert types. Each entry in a YAML file contains a unique name, a keyword list, and a sequence of diagnostic steps that an experienced engineer would typically follow. When an alert arrives, the store scores runbooks by keyword overlap against the alert text and injects the top three matches under a labelled prompt section. Keyword matching was chosen over vector search because Kubernetes alert names follow a standardised naming convention, making direct lexical overlap a reliable and low-maintenance retrieval strategy. In nine of the ten evaluation scenarios the runbook store contributed at least one match to the prompt.

The \textbf{vector memory module} retrieves semantically similar historical incidents using Sentence-BERT~\cite{reimers2019-sbert} embeddings stored in a FAISS~\cite{johnson2021-faiss} inner-product index over L2-normalised vectors. For two normalised vectors $\mathbf{a}$ and $\mathbf{b}$, the inner product equals the cosine similarity $\cos\theta = \mathbf{a} \cdot \mathbf{b}$, allowing FAISS's fast inner-product search to serve directly as a cosine nearest-neighbour lookup. The module returns the three nearest neighbours exceeding a cosine similarity threshold of 0.55. This threshold was set conservatively: Kubernetes alerts share heavy vocabulary across unrelated incident types, inflating baseline cosine similarity, and with only ten seeded post-mortems a misleading historical match would have a disproportionate effect on the agent's reasoning. Embeddings are computed over a combination of each incident's description and root cause fields, so incidents sharing an alert name but differing in underlying fault produce distinct representations.

The \textbf{context file layer} provides subsystem-specific background knowledge through a rules-based matching engine. Each context file covers one observability subsystem. For example, a NATS JetStream mirror drift context file contains background on how JetStream mirror synchronisation works, what the expected lag and ghost-key patterns look like in healthy versus degraded state, and which NATS CLI commands are most informative for a given symptom set. The engine activates a file when the incoming alert's label set matches the file's rules. In the evaluation, the NATS context file was the sole non-MCP retrieval source that contributed to Scenario~9 (NatsMirrorDrift), and practitioners credited it with enabling the quality of the NATS diagnosis.

The prompt builder assembles all retrieved context into the user message with clearly labelled sections. When a source yields no results, the prompt explicitly includes the section header followed by ``(none available)'', signalling to the model that the information was sought but not found. This deliberate choice encourages the model to mark those evidence dimensions as unknown rather than filling the gap with fabricated content.

\subsection{ReAct Loop \& Prompt Strategy}
\label{subsec:react}

The diagnostic core of ARGUS is an iterative investigation loop based on the ReAct paradigm~\cite{yao2023-react}, which interleaves reasoning, tool invocations, and observation. Rather than generating a diagnosis from a single LLM call over the enriched alert context, the agent iteratively decides which evidence to retrieve, executes the corresponding MCP tool call, observes the result, and determines whether to continue investigating or submit a final report. Figure~\ref{fig:e2e} shows the end-to-end processing flow, including MCP execution paths and the context-only fallback used when no server is reachable.

The system prompt adapts the eight-stage eARCO diagnostic framework~\cite{goel2025-earco}, instructing the agent to work through contextual identification, categorisation, symptom listing, historical review, environmental analysis, pattern analysis, root cause synthesis, and conclusion in a fixed sequence. Evidence-locking rules embedded in the prompt prohibit the model from making factual claims not supported by retrieved tool output: it must cite specific evidence artefacts by reference number, list any signals it was unable to retrieve, and write ``Unknown'' rather than infer a value when evidence is absent. The prompt also specifies an evidence hierarchy for different alert types, instructing the agent to prioritise Kubernetes events as primary evidence for pod and workload alerts, then fall back to pod logs including previous container output, and use Prometheus and Loki queries only when events and logs are inconclusive.

Three guardrail mechanisms constrain the loop. A \textbf{tool budget} caps the total number of MCP invocations per investigation, the calls per LLM turn, and the maximum iteration count. At the start of each iteration, a budget hint reporting the remaining call allowance is injected ephemerally into the system prompt. It is not persisted in the conversation history and therefore does not consume context window space across iterations. A \textbf{deduplication layer} detects when the LLM requests a call with the same name and arguments as one already executed in the same investigation and returns the cached result instead of re-executing the call. A \textbf{sufficiency gate} rejects the agent's final report submission if it lacks a minimum evidence set for the alert type. For pod and workload alerts, the gate requires pod or deployment state plus at least one confirming signal from logs, metrics, or events. For CrashLoopBackOff alerts, pod state and container logs are required. For resource and storage alerts, the gate requires the affected object's state plus at least one metric or log signal. If the gate rejects a submission, it returns feedback identifying the missing evidence categories and the agent continues the loop.

The investigation terminates under one of six conditions: (1) the LLM submits a final report and the sufficiency gate accepts it; (2) the LLM produces a text-only response with no tool calls; (3) the maximum iteration count is reached; (4) the total tool budget is exhausted; (5) the LLM API returns an error; or (6) no MCP tools are available. Each condition is recorded as a distinct stop reason in the investigation trace and surfaced in the Evidence Coverage block of the OperatorBrief.

Each connected MCP server is further protected by a per-server circuit breaker that opens after a configurable number of consecutive failures and recovers automatically after a timeout. Argument validation errors, such as a malformed PromQL expression, do not trip the circuit because they indicate a problem with the specific call rather than with the server. The underlying LLM is Claude Sonnet~4.6~\cite{anthropic2026sonnet}, selected for its strong agentic and tool-use performance, its 1M token context window (which accommodates large diagnostic payloads aggregating logs, metrics, and events), and its availability as a commercial API requiring no fine-tuning.

\subsection{OperatorBrief Output Format}
\label{subsec:operatorbrief}

\begin{figure}
\centering
\footnotesize
\begin{tikzpicture}[
  box/.style={draw, rounded corners, text width=0.74\linewidth, align=left,
              font=\footnotesize, inner sep=4pt},
  every node/.style={anchor=north},
]
\node[box] (s1) {\textbf{1.} Incident title $+$ severity};
\node[box, below=2pt of s1] (s2) {\textbf{2.} Plain-language explanation of the alert};
\node[box, below=2pt of s2] (s3) {\textbf{3.} Root cause statement $+$ causal explanation};
\node[box, below=2pt of s3] (s4) {\textbf{4.} Key evidence (source-attributed bullets)};
\node[box, below=2pt of s4] (s5) {\textbf{5.} Summary table of diagnostic findings};
\node[box, below=2pt of s5] (s6) {\textbf{6.} Numbered recommended fixes};
\node[box, below=2pt of s6] (s7) {\textbf{7.} Confidence label $+$ supporting note};
\node[box, fill=black!12, below=2pt of s7] (s8) {\textbf{8.} Evidence Coverage (collapsed by default)};
\draw[-{Latex[length=2mm]}] ($(s1.north west)+(-0.35,0)$) -- ($(s8.south west)+(-0.35,0)$)
  node[midway, rotate=90, anchor=south, font=\scriptsize\itshape] {situational awareness $\rightarrow$ supporting detail};
\end{tikzpicture}
\caption{Fixed section order of the OperatorBrief. The layout front-loads the most actionable content so an on-call engineer reads the incident framing and root cause first, with detailed evidence and coverage below. The shaded block is collapsed by default in the Slack message.}
\label{fig:brief-schematic}
\Description{A vertical stack of eight labeled boxes showing the fixed section order of an OperatorBrief, from incident title and severity at the top through plain-language explanation, root cause, key evidence, summary table, recommended fixes, confidence label, and a collapsed Evidence Coverage block at the bottom.}
\end{figure}
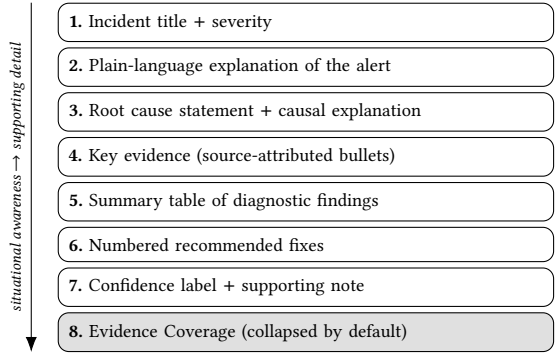

The agent's final output is an \emph{OperatorBrief}: a structured Slack message designed for rapid consumption under triage time pressure. The sections appear in a fixed order chosen to front-load the most actionable content, shown schematically in Figure~\ref{fig:brief-schematic}. The first two sections, an incident title with severity and a plain-language explanation of what the alert means, give the engineer immediate situational awareness. The root cause statement and causal explanation follow, each grounded in specific retrieved evidence. A bulleted key evidence list carries source-attribution prefixes, using distinct labels for alert payload, runbook, similar past incident, context file, and MCP tool output, so the engineer can see at a glance where each piece of evidence originated. A summary table collects the key diagnostic findings in scannable form. Numbered recommended fixes appear next. A confidence label and supporting note come before the final section: the Evidence Coverage block, which lists the runbooks consulted, similar incidents retrieved, context files activated, and a per-server breakdown of MCP tool calls by success, failure, and skip count.

Since LLMs are known to not be well-calibrated~\cite{Spiess2025calibration,Virk2025calibration} (i.e., they struggle with correctly determining their own confidence), the confidence label is computed deterministically from the investigation trace rather than generated by the model. Table~\ref{tab:confidence} shows the complete rule set. It is a function of the MCP success ratio $R = S / T$ and the count $N \in \{0,1,2,3\}$ of non-MCP retrieval source types that contributed to the prompt. Higher confidence requires both reliable tool execution and corroboration from multiple retrieval sources. The deterministic design replaced a self-reported model confidence score used in a prototype version, which was found to be uncalibrated and inconsistent across runs of the same scenario.

\begin{table}
\centering
\caption{Rule-based confidence label assignment from the MCP success ratio $R = S/T$ and non-MCP source count $N$. $T$ is the total number of MCP tool calls made during the investigation and $S$ the number that succeeded, so $R = S/T$ is the MCP success ratio (with $R$ undefined when $T = 0$, handled by the first two rows). $N \in \{0,1,2,3\}$ is the number of non-MCP retrieval sources (runbook store, similar past incidents, context files) that contributed to the prompt.}
\label{tab:confidence}
\footnotesize
\begin{tabularx}{\linewidth}{@{}X l@{}}
\toprule
\textbf{Condition} & \textbf{Label} \\
\midrule
$T = 0$ or $R = 0$, with $N = 0$ & Very Low \\
$T = 0$ or $R = 0$, with $N > 0$ & Low \\
$0 < R < 0.6$, $N = 0$ & Low \\
$0.6 \le R < 1.0$, $N = 0$ & Medium \\
$0 < R < 0.6$, $N > 0$ & Medium \\
$0.6 \le R < 1.0$, $N > 0$ & High \\
$R = 1.0$, $N < 2$ & High \\
$R = 1.0$, $N \ge 2$ & Very High \\
\bottomrule
\end{tabularx}
\end{table}

\begin{figure*}
    \centering
    \begin{tikzpicture}
        \node[anchor=south west, inner sep=0] (bimg) {\includegraphics[clip, viewport=0 1227.25 905 2454.5, width=0.49\linewidth]{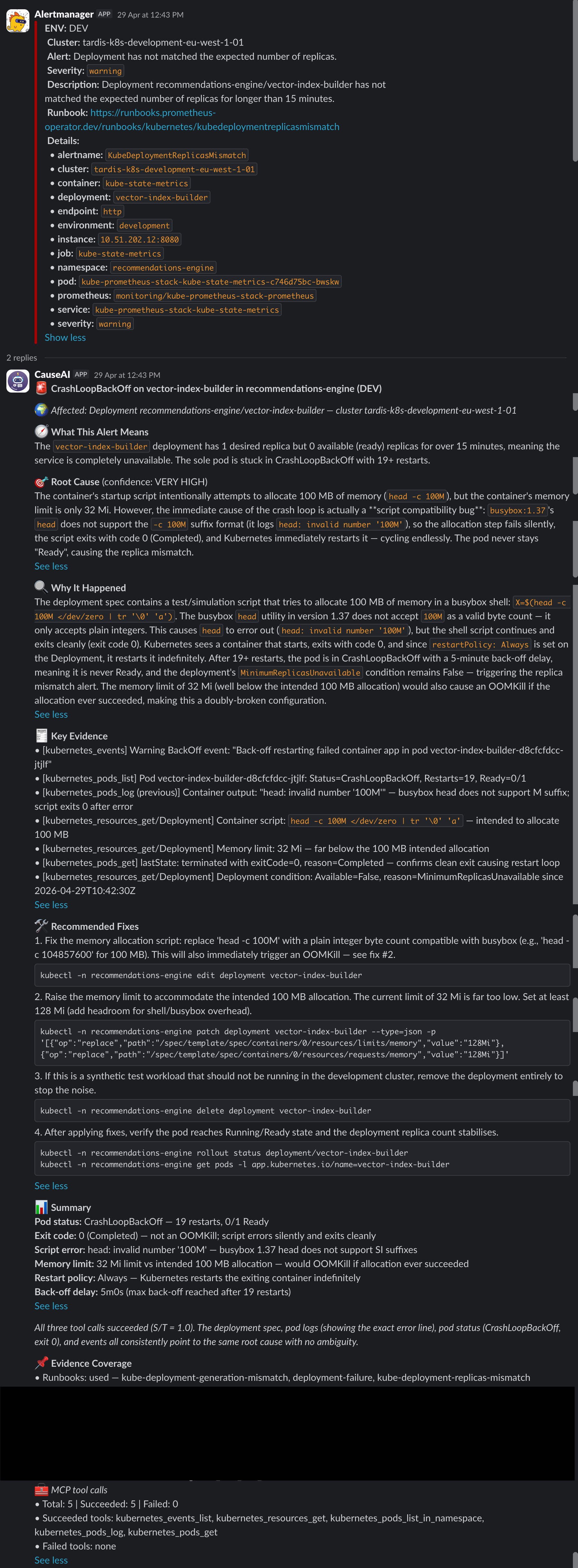}};
        \fill[black]
          ($($(bimg.north west)!0.045!(bimg.north east)$)!0.467!($(bimg.south west)!0.045!(bimg.south east)$)$)
          rectangle
          ($($(bimg.north west)!0.18!(bimg.north east)$)!0.486!($(bimg.south west)!0.18!(bimg.south east)$)$);
    \end{tikzpicture}\hfill
    \includegraphics[clip, viewport=0 0 905 1227.25, width=0.49\linewidth]{cycle3-output.png}
    \caption{Example OperatorBrief delivered as a Slack thread reply, shown in full and split into two columns for space (left: the incident framing, root cause with confidence label, and causal explanation; right: the source-attributed key evidence, numbered recommended fixes, summary, and the Evidence Coverage and per-server MCP tool-call breakdown, shown expanded, at the bottom). Diagnostic text and the local tool name are redacted not to reveal the industrial partner.}
    \label{fig:operatorbrief}
    \Description{Screenshot of an OperatorBrief in Slack showing the incident title and severity, an explanation of the alert, a root cause with a confidence label, a causal explanation, key evidence with source-attribution prefixes, numbered recommended fixes, a summary, an Evidence Coverage block, and a per-server MCP tool-call breakdown.}
\end{figure*}

All technical content in the OperatorBrief, including resource names, log excerpts, and recommended commands, is formatted in Slack markdown code blocks to distinguish it from surrounding prose. The Evidence Coverage block is delivered with the Slack block-kit \texttt{expand=false} attribute so it appears collapsed by default and the engineer expands it in place when investigating the completeness of the agent's evidence gathering. Figure~\ref{fig:operatorbrief} shows an example OperatorBrief from the evaluation.

\section{Evaluation Methodology}
\label{sec:methodology}

\begin{table*}[t]
\centering
\footnotesize
\caption{The six RQ3 interview participants. Numbering is randomised and does not preserve interview order. ``Inc./mo'' is the self-reported approximate monthly incident count.}
\label{tab:participants}
\begin{tabular}{@{}c p{3.6cm} c c c c p{3.6cm}@{}}
\toprule
\textbf{\#} & \textbf{Role} & \textbf{Exp.\ (yr)} & \textbf{Tenure} & \textbf{Inc./mo} & \textbf{Kubernetes use} & \textbf{Prior AI use in incidents} \\
\midrule
1 & SRE, incident-response focus & 6 & 4+\,yr & $\sim$30 & Daily & Daily LLM coding agent \\
2 & Cloud-ops / platform engineer & $\sim$18 & 4+\,yr & $\sim$12 & Daily & Daily LLM coding agent \\
3 & Solutions architect / backend & 15 & 4+\,yr & $<$5 & Seldom & None (declines AI-written code) \\
4 & Cloud \& information security & $\sim$4 & $\sim$2\,yr & $\sim$2 & Daily & None (writing/research only) \\
5 & Platform engineer & 15 & 3\,mo & 0--1 & Daily & Code-side only \\
6 & Tech lead / engineering manager & 10 & 3+\,yr & 0 (aftermath) & Daily (observer) & None \\
\bottomrule
\end{tabular}
\end{table*}

ARGUS was developed following a Design Science Research (DSR) methodology~\cite{hevner2004design} across three iteration cycles. Cycles~I and~II focused on design, implementation, and formative evaluation. Cycle~I validated the reasoning pipeline, prompt architecture, and Slack integration using static mocked tool responses. Cycle~II introduced live MCP integration against a real Kubernetes cluster and refined the OperatorBrief format based on practitioner feedback. Cycle~III conducted the summative evaluation reported here. This evaluation is preliminary: conducted at a single industrial site with a small set of scenarios and participants, it is intended to establish initial feasibility and surface design implications for MCP-grounded RCA agents rather than to provide generalisable performance guarantees.

The evaluation is structured around three research questions:
\begin{description}
  \item[\textbf{RQ1:}] To what extent can an MCP-grounded agent retrieve and compile the incident-relevant evidence needed for RCA?
  \item[\textbf{RQ2:}] How do the resulting RCA summaries perform with respect to correctness, evidence backing, and operational usefulness?
  \item[\textbf{RQ3:}] How do on-call engineers perceive the agent's usefulness, convenience, and impact on how quickly they reach a first actionable hypothesis?
\end{description}

\textbf{RQ1} is addressed through controlled fault injection in a dedicated test Kubernetes cluster mirroring the industrial partner's development environment. Ten scenarios cover a representative range of incident types. Pod-level failures include CrashLoopBackOff, OOMKilled, Image\-Pull\-Back\-Off, and a readiness probe hardcoded to fail. Workload-level failures include three Kube\-Deploy\-ment\-Replicas\-Mismatch alerts from three distinct root causes, HPA saturation, and a failed CronJob with NATS JetStream involvement. One node-level failure (Kube\-Node\-Not\-Ready) and one storage failure (Persistent\-Volume\-Claim exhaustion) complete the set. Three scenarios deliberately share the same alert name but inject different underlying faults, testing whether the agent disambiguates through retrieved evidence rather than alert-name pattern matching. For each scenario, an expected evidence set, expressed as the list of MCP tool categories an experienced on-call engineer would consult to form a confident hypothesis, was defined and fixed before any agent run was executed to avoid post-hoc rationalisation. Coverage is classified as \emph{Full} when all expected tool categories were retrieved and as \emph{Partial} when at least one expected category was missed due to a failure or was never invoked.

\textbf{RQ2} is addressed by evaluating the ten RCA summaries against a structured rubric on three dimensions: \emph{Correctness} (does the root cause hypothesis name the specific fault mechanism, e.g.\ the exact image tag or the exact misconfigured field?), \emph{Evidence Backing} (is every factual claim in the root cause statement supported by a retrieved artefact cited in the output?), and \emph{Actionability} (are the recommended fixes concrete enough to execute as written, with the target namespace, resource name, and value specified?). Each dimension is scored on a three-point scale (Fully met, Partially met, Not met) using anchors fixed before any scoring began. Scores were assigned independently by two raters, and disagreements were resolved by a practising on-call engineer at the industrial partner as a domain-informed third-party adjudicator, following the procedure of Chinh et al.~\cite{Chinh2019WaysOQ}.

\textbf{RQ3} is addressed through semi-structured interviews with six on-call engineers at the industrial partner. Participants were selected using purposive sampling~\cite{Palinkas2015Purposeful}, a non-probability strategy well suited to predominantly qualitative software engineering research~\cite{Baltes2022sampling}, to achieve variation across Kubernetes experience (ranging from a three-month newcomer to an eighteen-year industry veteran), AI tooling familiarity (from no prior use to daily use of LLM-based coding agents), and seniority. Each interview presented a selection of ARGUS-generated OperatorBriefs from the fault injection scenarios, and participants were asked to evaluate usefulness, clarity, and perceived impact on time-to-hypothesis. Interviews were audio-recorded with consent, transcribed, and analysed using thematic analysis following Braun and Clarke~\cite{braun2006thematic}, with codes derived inductively from line-level transcript annotations. Table~\ref{tab:participants} summarises our interviewees. Participant demographics deliberately span the full range of incident exposure, from an SRE handling roughly thirty incidents a month to a newcomer of three months and a tech lead who now engages only with incident aftermath.

\section{Results}
\label{sec:results}

We now discuss the results of our preliminary evaluation.

\subsection{RQ1: Evidence Retrieval}
\label{subsec:rq1}

Across the ten fault injection scenarios, ARGUS issued 82 MCP tool invocations, of which 75 succeeded, giving an aggregate success ratio of $S/T = 0.91$. Eight scenarios achieved full coverage of the expected evidence set. Two scenarios were classified as partial: KubeNodeNotReady (Scenario~5), in which four node-scoped calls were denied by RBAC, and NatsMirrorDrift (Scenario~9), in which two stream-info calls failed due to missing credentials for development NATS accounts. Despite partial coverage, ARGUS produced an OperatorBrief naming the correct root cause in all ten scenarios, recovering through alternative tool paths in both partial-coverage cases. Table~\ref{tab:rq1} summarises coverage per scenario.

\begin{table}[h!]
\centering
\caption{Evidence retrieval coverage per scenario. S/T: successful over total MCP calls. Non-MCP sources: R=runbook, S=similar incident, C=context file. A letter is omitted when the source produced no match.}
\label{tab:rq1}
\footnotesize
\setlength{\tabcolsep}{3pt}
\begin{tabularx}{\linewidth}{@{}c p{3.0cm} c c l@{}}
\toprule
\# & Scenario & S/T & Non-MCP & Coverage \\
\midrule
1 & HighErrorRate & 20/20 & R,S,C & Full \\
2 & ReplicasMismatch (image tag) & 2/2 & R,S,C & Full \\
3 & HpaMaxedOut & 4/4 & R,S,C & Full \\
4 & JobFailed (NATS mirror) & 10/10 & R,S,C & Full \\
5 & NodeNotReady & 7/11 & R,S,C & \textbf{Partial} \\
6 & PVFillingUp & 7/8 & R,S,C & Full \\
7 & PodCrashLooping & 6/6 & R,S,C & Full \\
8 & ReplicasMismatch (probe) & 3/3 & R,S,C & Full \\
9 & NatsMirrorDrift & 11/13 & C & \textbf{Partial} \\
10 & ReplicasMismatch (busybox) & 5/5 & R,S,C & Full \\
\midrule
\textbf{Total} & & \textbf{75/82} & & \textbf{8/10 Full} \\
\bottomrule
\end{tabularx}
\end{table}

The three KubeDeploymentReplicasMismatch scenarios (2, 8, 10) fire the same alert from three distinct underlying faults: a non-existent container image tag, a readiness probe hardcoded to \texttt{exit 1}, and a busybox \texttt{head~-c 100M} invocation that fails because busybox does not accept the \texttt{M} size suffix. In all three cases the agent retrieved the deployment specification and recent cluster events before proposing a hypothesis, and in each case the hypothesis matched the injected fault. The diagnosis was driven by the retrieved evidence rather than by alert-name pattern matching.

The seven failed calls fall into three distinct infrastructure failure modes. The first is RBAC restriction on node-level resources: in Scenario~5, four calls targeting node objects were denied with a permission error because the Kubernetes MCP service account lacked cluster-scoped \texttt{get} permission on the \texttt{nodes} resource. The agent compensated by using cluster events and kubelet metrics retrieved via Prometheus, which together were sufficient to identify the loss of kubelet heartbeat as the cause. The second is NATS account credential gaps: in Scenario~9, two \texttt{nats\_stream\_info} calls failed because the NATS MCP server held no credentials for the development-environment accounts owning the affected JetStream KV buckets. The agent recovered via \texttt{nats\_kv\_info} and \texttt{nats\_stream\_state}, which use a different credential path. The third is an API group denial: a single \texttt{StorageClass} lookup in Scenario~6 was denied by the RBAC policy on the \texttt{storage.k8s.io} API group, but this call was incidental because the root cause had already been established from the pod logs and deployment spec. In all three modes, the failures were surfaced explicitly in the Evidence Coverage block rather than silently omitted.

\subsection{RQ2: Summary Quality}
\label{subsec:rq2}

Table~\ref{tab:rq2} reports the consensus rubric scores. Pre-consensus inter-rater agreement was 83.3\% exact across the 30 cells (10 scenarios $\times$ 3 dimensions), with no disagreement exceeding one rubric level. Cohen's $\kappa$ over the 30 cells was 0.36 (fair agreement on the Landis and Koch scale), but this reflects the heavily skewed label distribution (85\% of cells received Fully met) rather than rater inconsistency. The prevalence-robust Gwet's AC1~\cite{gwet2008}, designed for skewed distribution, was 0.81 (substantial agreement). The adjudicator changed nine cells: the five first-pass disagreement cells and four cells where both raters had agreed on Fully met, confirming that domain-expert review surfaces issues that shared-background raters miss even when they agree.

\begin{table}[h!]
\centering
\caption{Consensus rubric scores. F=Fully met, P=Partially met, N=Not met. $\dagger$ marks cells the adjudicator downgraded relative to both raters' first-pass agreement.}
\label{tab:rq2}
\footnotesize
\setlength{\tabcolsep}{4pt}
\begin{tabularx}{\linewidth}{@{}c p{3.0cm} c c c@{}}
\toprule
\# & Scenario & Corr. & Evid. & Act. \\
\midrule
1 & HighErrorRate & P & P$^{\dagger}$ & N$^{\dagger}$ \\
2 & ReplicasMismatch (image) & F & F & N$^{\dagger}$ \\
3 & HpaMaxedOut & F & F & F \\
4 & JobFailed (NATS) & F & F & F \\
5 & NodeNotReady & F & P & F \\
6 & PVFillingUp & F & F & F \\
7 & PodCrashLooping & P$^{\dagger}$ & F & P$^{\dagger}$ \\
8 & ReplicasMismatch (probe) & F & F & F \\
9 & NatsMirrorDrift & F & P & F \\
10 & ReplicasMismatch (busybox) & F & F & F \\
\midrule
\textbf{Total F/P/N} & & 8/2/0 & 7/3/0 & 7/1/2 \\
\bottomrule
\end{tabularx}
\end{table}

The two Not~met cells on Actionability are the most informative entries in the matrix. Scenario~1 was downgraded because the recommended actions enumerated high-level connection checks without scoped commands, while the diagnosis itself identified genuine application issues those checks would not address. Scenario~2 was downgraded because the agent proposed an \texttt{aws ecr describe-images} command for an image sourced from Docker Hub rather than ECR: the prescription named the wrong registry entirely. These two failures sit at opposite ends of a spectrum, one too vague and one factually incorrect, but both concern the prescriptive part of the output rather than the diagnostic part. No cell on Correctness or Evidence Backing received Not~met.

Four further cells were scored Fully~met by both raters and downgraded only once the adjudicator applied operational judgement (marked $\dagger$ in Table~\ref{tab:rq2}), and each exposes an output that reads as correct but that an experienced responder would not act on. In Scenario~1, the fix told the engineer to inspect Vault connectivity, yet no Vault signal appeared anywhere in the retrieved evidence. In Scenario~2, the suggested ECR command pointed at a registry that does not even host the affected image. Scenario~7 failed on two axes at once: the diagnosis named the failing pod but missed a ConfigMap reference in the pod manifest, and the matching fix omitted the configuration-level check its own commands needed before they would run. The common thread is a boundary the rubric cannot see on its own, the gap between what the output claims and what an engineer would actually have to do.

\subsection{RQ3: Practitioner Perception}
\label{subsec:rq3}

\begin{figure}[h!]
\centering
\resizebox{\linewidth}{!}{%
\begin{tikzpicture}[
  font=\footnotesize,
  root/.style={draw, rounded corners, fill=black!14, align=center, inner sep=4pt},
  cat/.style={draw, rounded corners, fill=black!7, align=center, inner sep=4pt, minimum height=6mm},
  leaf/.style={draw, rounded corners, align=left, inner sep=3pt, minimum height=5.5mm, anchor=west},
  leafall/.style={leaf, fill=green!15, draw=green!50!black, thick},
  bubble/.style={draw, rounded corners=4pt, fill=blue!12, font=\scriptsize, inner sep=2.5pt, anchor=east},
  bubbleall/.style={bubble, fill=green!30, draw=green!50!black},
  link/.style={draw, black!55},
]
\node[root] (root) at (0,-3.325) {Practitioner\\perception};
\node[cat] (c1) at (2.5,-0.95)  {Usefulness};
\node[cat] (c2) at (2.5,-3.325) {Convenience};
\node[cat] (c3) at (2.5,-5.7)   {Time-to-\\hypothesis};
\node[leafall] (t1) at (4.7, 0)  {\textbf{T1} Diagnosis trusted, fixes doubted};
\node[leaf] (t2) at (4.7,-0.95)  {\textbf{T2} Bounded by MCP coverage};
\node[leaf] (t3) at (4.7,-1.9)   {\textbf{T3} Bridges specialist gaps};
\node[leaf] (t4) at (4.7,-2.85)  {\textbf{T4} Wants shorter output};
\node[leafall] (t5) at (4.7,-3.8) {\textbf{T5} Confidence label: mixed};
\node[leaf] (t6) at (4.7,-4.75)  {\textbf{T6} Time savings domain-specific};
\node[leaf] (t7) at (4.7,-5.7)   {\textbf{T7} Decision point shifts earlier};
\node[leafall] (t8) at (4.7,-6.65) {\textbf{T8} Trust is built, not granted};
\node[bubbleall] (b1) at (9.9, 0)     {All};
\node[bubble] (b2) at (9.9,-0.95)     {P1,P2,P3,P6};
\node[bubble] (b3) at (9.9,-1.9)      {P3,P4,P5,P6};
\node[bubble] (b4) at (9.9,-2.85)     {P2,P3,P4,P5};
\node[bubbleall] (b5) at (9.9,-3.8)   {All};
\node[bubble] (b6) at (9.9,-4.75)     {P1,P2,P3,P5};
\node[bubble] (b7) at (9.9,-5.7)      {P1,P2,P5,P6};
\node[bubbleall] (b8) at (9.9,-6.65)  {All};
\draw[link] (root.east) -- (c1.west);
\draw[link] (root.east) -- (c2.west);
\draw[link] (root.east) -- (c3.west);
\draw[link] (c1.east) -- (t1.west);
\draw[link] (c1.east) -- (t2.west);
\draw[link] (c1.east) -- (t3.west);
\draw[link] (c2.east) -- (t4.west);
\draw[link] (c2.east) -- (t5.west);
\draw[link] (c3.east) -- (t6.west);
\draw[link] (c3.east) -- (t7.west);
\draw[link] (c3.east) -- (t8.west);
\end{tikzpicture}%
}
\caption{Themes from the six practitioner interviews, organised as a taxonomy over the three interview sub-dimensions. Each theme node carries a bubble listing the participants who contributed to it. ``All'' denotes convergence across all six participants. Full-consensus themes (T1, T5, T8) are highlighted in green.}
\label{fig:rq3-themes}
\Description{A horizontal taxonomy tree rooted at ``Practitioner perception'' branching into three sub-dimensions (Usefulness, Convenience, Time-to-hypothesis), each branching into its themes T1 to T8. Each theme node has an attached bubble listing the contributing participants.}
\end{figure}
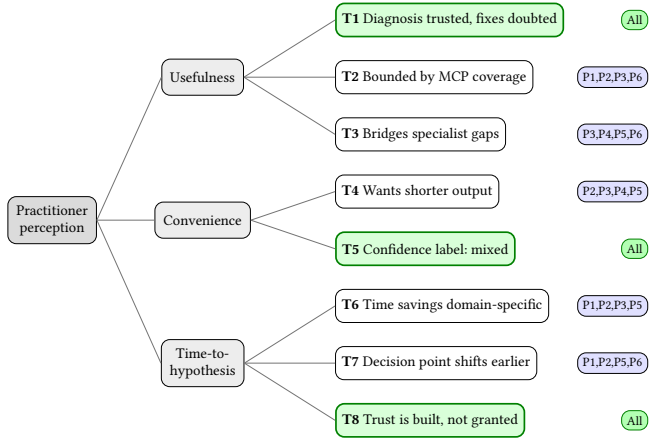

Thematic analysis of the six interviews yielded eight themes across three sub-dimensions, summarised as a taxonomy in Figure~\ref{fig:rq3-themes}.

Theme~\textbf{T1} is the core finding. All six participants read the diagnostic core of the OperatorBrief to form their first hypothesis, and all six singled out the Recommended Fixes block as the part they questioned. One put the posture directly: \textit{``If it did not include recommended fixes, I would fully trust it, because it would only be telling me what is happening, not what to do.''} Participants converged on read-only-by-default, with capability expansion only after sustained exposure to correct output.

Theme~\textbf{T2} surfaced in four interviews: every failure participants cited mapped to a boundary of the agent's MCP access rather than a defect in its reasoning. One described a real \texttt{KubeJobFailed} incident in which ARGUS correctly identified that the image could not be pulled, but the actual root cause was an ECR lifecycle policy that had deleted it, invisible because the agent had no AWS MCP server. The output was incomplete rather than wrong, yet the incompleteness was not evident from the brief itself. Participants independently named extending MCP coverage to the AWS control plane as the most valuable next step.

Theme~\textbf{T3} casts ARGUS as a knowledge equaliser at the boundary of team expertise. The three-month newcomer described the NatsMirrorDrift brief as giving enough grounding to understand the misconfiguration and identify the next step on a subsystem that was entirely unfamiliar, whereas a senior engineer noted that where they already pattern-match an alert from memory the tool adds little. Theme~\textbf{T6} is the time equivalent of this gradient: twenty to thirty minutes saved on unfamiliar NATS scenarios versus two to five minutes on routine Kubernetes ones.

Theme~\textbf{T4} reflects a convenience concern raised by four of the six participants: the current OperatorBrief format, while structured, is too dense for consumption under active triage stress. Three specific requests recurred independently. The first is a brief summary at the very top of the message, a one or two sentence encapsulation the engineer can read in ten seconds before deciding whether to continue. One participant described discovering in the interview that they routinely scroll past the initial sections to read the summary block first, then returns to the top, suggesting the current ordering does not match their actual reading pattern. The second is stronger visual hierarchy: the affected namespace and deployment name are rendered in italics but participants wanted them bolded or hyperlinked directly to the relevant Grafana query or pod object. The third is reduced internal repetition between the root cause statement and the causal explanation, which several participants found to restate the same information in different words without adding signal. The two participants who did not raise these concerns interact with the output either at very high frequency (where familiarity reduces scanning cost) or in a post-incident review context (where time pressure is absent).

Theme~\textbf{T5} reports a divided reception for the deterministic confidence label. Two participants read it as a useful calibration signal that tracks the gap between full-coverage scenarios (Very High) and those with credential or RBAC holes (High or Medium), while three either ignored it under stress or saw it as harmful. The sharpest concern, from a participant with high-stakes operational experience, was that an early high-confidence label narrows the option space: \textit{``If we present that too early with too high a confidence, we guide the team into the wrong direction.''} Several proposed an asymmetric rule: the label may be lowered on explicit evidence gaps but never raised above what the evidence supports. The split is at root a construct gap: the label reports evidence quality, whereas participants who dismissed it were implicitly asking a different question, whether the diagnosis itself was correct.

Theme~\textbf{T7} is subtler: rather than speeding an existing investigation, the agent moves the decide-to-engage step earlier. One participant skipped the habitual Kubernetes check and went straight to the AWS console once the brief named the missing image, and another could assess an off-hours alert and decide whether to escalate without opening a laptop, a saving in disrupted personal time rather than investigation time. Theme~\textbf{T8} is the trust trajectory: no participant claimed unconditional trust, and all described accumulating evidence from correct outputs before granting the agent further capability. The trajectory they endorsed was explicit and graduated: whitelisted commands first, then changes that can only restrict rather than widen access, and approve-before-execute workflows, with full autonomy proposed by none of the participants.

\subsection{Discussion}
\label{subsec:discussion}

We now discuss the main lessons learned from our study. Afterwards, we briefly summarize the limitations of ARGUS in its current version that have emerged in our evaluation.

\textbf{Diagnosis is perceived as reliable, prescription is not.} Across all three methods, ARGUS is right about what went wrong, but suggestions what to do next are perceived as unreliable. The tool named the correct root cause in every scenario and drew universal trust in its diagnostic output, yet the RQ2 consensus places its only Not~met cells on Actionability and RQ3 records universal scepticism toward the Recommended Fixes block. This implies that command-proposing capability should be a separately authorised function rather than a default. Additionally, these results carry larger implications for future fully autonomous agentic systems for DevOps, incident management, or autonomous coding.

\textbf{The dominant limitation of the tool is the MCP server inventory, not the model.} All seven failed tool calls were infra\-structure-level issues, never reasoning errors such as invalid PromQL, a hallucinated tool, or a misread observation. Reporting tool-reach and reasoning quality on separate axes is what made this visible, whereas a single end-to-end accuracy score would have shown 10/10 and hidden the real limit. Extending diagnostic capability therefore means extending MCP coverage rather than changing the model or prompt. This observation is similar to findings by OpenDerisk~\cite{di2025-openderisk} at production scale.

\textbf{A tool like ARGUS helps most at the edge of a team's expertise.} Its largest gains appeared on unfamiliar subsystems, where it gave a newcomer enough grounding to take the next step and saved twenty to thirty minutes on NATS scenarios, while adding little for a senior engineer already pattern-matching a familiar alert. Teams that frame the value primarily as a speed-up for their most experienced responders are likely to be disappointed, and should instead treat it as a leveller of specialist knowledge gaps.

\textbf{Limitations.}
Despite its promises, our evaluation has also shown some limitations in the current version of ARGUS. Its diagnostic reach is bounded by its MCP server inventory: gaps such as the absent AWS control plane leave evidence holes that the OperatorBrief does not itself flag. It is read-only and advisory, so it proposes but never executes fixes, and those recommended fixes are the least reliable part of its output and require human verification before use. Finally, it is a single ReAct agent scoped to Kubernetes RCA, without causal-graph construction or multi-agent orchestration.

\subsection{Threats to Validity}
\label{subsec:threats}

\textbf{Internal validity.} Each scenario was run only once, so the impact of non-determinism of the underlying LLM cannot be quantified. However, given the general consistency of our results we believe the impact of run-to-run variation to be relatively small. Further, the expected-evidence sets for RQ1 and the rubric anchors for RQ2 were defined by the researchers, which risks confirmation bias. To counter this, both were fixed before any agent run, and disputed scores were settled by a practising on-call engineer acting as a domain-informed adjudicator. Finally, the interviews carry a social-desirability risk, since participants assessed a tool developed within their own organisation, which may inflate positive sentiment. To limit this, the interview guide instructed the interviewer to stay neutral and neither affirm nor defend the tool, and the spread of critical feedback, on the recommended fixes, the output length, and the confidence label, indicates that participants did voice disagreement. A construct concern also applies to RQ2, where the Actionability anchor scores whether a fix is executable as written, a property that does not match how engineers actually consume fixes, since they verify each command before running it.

\textbf{External validity.} The key limitation of our evaluation is that we were only able to assess ARGUS in a single-case study at one industrial site, on one test cluster, with six engineers from one on-call team. Hence, coverage figures and interview themes reflect one organisation's RBAC configuration, observability stack, and operational culture. The ten scenarios emphasise configuration-level faults rather than the resource-pressure faults common in prior benchmarks, so results may differ under CPU or memory saturation. The agent was driven by a single commercial LLM, and a different model could shift both the diagnostic and the prescriptive results. The interview sample is small ($n = 6$), so themes supported by only a subset of participants (Figure~\ref{fig:rq3-themes}) warrant more caution before being generalised.

\section{Conclusion}
\label{sec:conclusion}

We presented ARGUS, a Slack-integrated, read-only RCA assistant that grounds a commercial LLM in live Kubernetes observability data through standardised MCP servers. A three-method preliminary evaluation combining ten fault injection scenarios, a rubric with domain-expert adjudication, and six practitioner interviews established two principal findings. First, ARGUS named the correct root cause in all ten scenarios with an aggregate MCP tool success ratio of 0.91, demonstrating that protocol-mediated grounding produces reliable evidence retrieval without model fine-tuning or bespoke data access layers. Second, ARGUS reliably identifies what went wrong, but its recommended fixes are the least reliable and least trusted part of its output. 

For future work, three directions stand out. First, extend MCP coverage to the AWS control plane and beyond, and have the agent flag what it could not observe (T2). Second, close the diagnostic/prescriptive gap by grounding recommended fixes in verified live state, gating them behind graduated, approve-before-execute autonomy (T8), and making the confidence label calibrated and able only to lower itself on evidence gaps (T5). Third, strengthen the evidence base with repeated runs, control-plane and cross-cloud scenarios, and a multi-organisation, multi-model replication.

\section*{Acknowledgments}
This work received financial support from the Swedish Research Council VR under grant number 2025-04346 (Fast-LLM, Can Large Language Models Synthesise Efficient Code?). The authors acknowledge the controlled use of generative AI (Claude) for implementation support, data analysis, and writing assistance.

\bibliographystyle{ACM-Reference-Format}
\bibliography{references}

\end{document}